\documentclass[preprint2]{aastex}

\slugcomment{Not to appear in Nonlearned J., 45.}

\shorttitle{Dynamical Heating Induced by Dwarf Planets}
\shortauthors{Mu\~noz-Guti\'errez et al.}

\begin{document}


\title{Dynamical Heating Induced by Dwarf Planets\\ 
on Cold Kuiper Belt-like Debris Disks}


\author{M. A. Mu\~noz-Guti\'errez, B. Pichardo}
\affil{Instituto de Astronom\'ia, Universidad Nacional Aut\'onoma de M\'exico, Apdo. postal 70-264 Ciudad Universitaria, M\'exico}
\email{mmunoz@astro.unam.mx}

\author{M. Reyes-Ruiz}
\affil{Instituto de Astronom\'ia, Universidad Nacional Aut\'onoma de M\'exico, Apdo. postal 877, 22800 Ensenada, M\'exico}

\and

\author{A. Peimbert}
\affil{Instituto de Astronom\'ia, Universidad Nacional Aut\'onoma de M\'exico, Apdo. postal 70-264 Ciudad Universitaria, M\'exico}


\begin{abstract}

With the use of long-term numerical simulations, we study the
evolution and orbital behavior of cometary nuclei in cold Kuiper
belt-like debris disks under the gravitational influence of dwarf
planets (DPs); we carry out these simulations with and without the
presence of a Neptune-like giant planet. This exploratory study
shows that in the absence of a giant planet, 10 DPs are enough to
induce strong radial and vertical heating on the orbits of belt
particles. On the other hand, the presence of a giant planet close to
the debris disk, acts as a stability agent reducing the radial and
vertical heating. With enough DPs, even in the presence of a
Neptune-like giant planet some radial heating remains; this heating
grows steadily, re-filling resonances otherwise empty of cometary
nuclei. 
Specifically for the solar system, this secular process seems to be able 
to provide material that, through resonant chaotic diffusion, increase 
the rate of new
comets spiraling into the inner planetary system, but only if more 
than the $\sim10$ known DP sized objects exist in the 
trans-Neptunian region.

\end{abstract}

\keywords{planets and satellites: dynamical evolution and stability --- Kuiper belt: general --- methods: numerical } 

\section{Introduction}

One of the characteristics of evolved planetary systems is the
prolonged presence of the remnants of stellar and planetary formation,
ranging in size from dust grains to cometary nuclei to DPs. This
material, located beyond the region where planets rapidly ``clean-up''
their vicinity, is known as a debris disk \citep[for a review see][and
references therein]{WARAA08,KBOD08}. In our solar system the present
day remnants in this region constitute the ``Kuiper Belt''
(KB). Although the lifetime of debris disks depends on diverse
factors, such as the stellar mass and neighboring environment, the
majority of 100 Myr old stars have observational features consistent
with the presence of debris disks and even a few 10 Gyr old stars show
evidence of having debris disks \citep{DD03,GHW05}.

The first discovered extrasolar debris disk was the one of Vega,
detected by its infrared (IR) excess with the IRAS satellite
\citep{Aumann84}. The IR excess is believed to be produced by belts of
dust particles originating from a steady collisional cascade
\citep{Muller10}; for the case of Vega, this belt is located at
$~\sim100$ AU from the star. The study of extrasolar debris disks is
relevant in several aspects to the understanding of the planetary
system formation process; moreover, debris disks have been employed to
determine the presence of planets in extrasolar planetary systems
\citep{ZS04}.

On the other hand, DPs have an important role on the dynamics of
primigenious planetary disks as the initiators of collisional cascades
once they reach $\sim1000$ to 3000 km size; they stir the orbits of
residual icy planetesimals, increasing collisions; these collisions
are responsible for both grounding some icy-planetesimals to dust, as
well as creating some super-Earth sized cores \citep{KB04,KB15}. Also,
massive planets in evolved debris disks are able to produce gaps and
dust outflows \citep{Moro05}.

In the specific case of the KB, recent studies show that a number of
its dynamical components can be explained with a migrating Neptune
\citep[e.g.][]{Malhotra93,Levison03,CL07,Morbidelli08,Nesvorny15}.
Indeed, all populations in the KB conserve evidences of their close
interaction with the giant, except probably for the classical KB
(CKB). The CKB has been defined as a bimodal orbital distribution: the
hot (inclinations $i>5^{\rm o}$) and cold ($i<5^{\rm o}$) components
\citep{Brown01}. However, some mixing between both populations seem to
have taken place \citep{Morbidelli08,Volk11,Petit11}.

The most accepted scenario to explain the coexistence of both hot and
cold populations \citep{BBF11,Wolff12,Nesvorny15} involves the action
of a migrating Neptune, going outwards launching lots of planetesimals
to form the hot population; the cold disk bodies, starting beyond 40
AU, simply kept their primordial orbits mostly unaffected by Neptune
that stopped migrating at some point in the evolution of the early
solar system when the disk material grew scarce \citep{GM2004}.

Regarding the largest bodies of the power spectra on debris disks, the
only examples we know are the KB objects (KBOs) with radii between 400
and 1200 km, a few of which have only recently been discovered
\citep{Brown05}. Extrapolation of the size distribution of smaller
KBOs has sometimes been used to attempt to estimate the numbers of
such larger objects \citep[i.e.][]{Bernstein04}, but estimations are
still inconclusive.

Regardless of their number, it is usually believed DPs to have only a
small influence on the evolution of debris disks in
general. \citet{FJ80} presents a first approximation where he attaches
importance to massive objects, of up to $1.7\times10^{-4}{\rm
  M}_{\oplus}$, in a very massive KB disk (about 9 ${\rm
  M}_{\oplus}$), finding that, in the presence of thousands of
Ceres-like objects, direct encounters of cometary nuclei with larger
bodies could lead to scatter of comets, sending them to the inner
planetary region, in this way possibly maintaining a steady influx of
short-period comets. Current estimates of the mass and composition of
the KB rule out this possibility as the main driver to produce the
observed population of short-period comets. The infall inrate of
comets on planetary systems might be of great importance in terms of
habitability for example: it is believed that a large fraction of the
water in the primeval Earth came from comets and asteroids
\citep{ABB15}; also, at some later point it becomes necessary, for
long-term evolution of life, to have a reduced cometary infall
rate. However at present, other than the KB, we are not able to
observe such details on other debris disks.

In this work we produce an exploratory study, that helps to better
understand the dynamical effects of DPs on cold Kuiper belt-like
debris disks (KBLDD) with and without the influence of a Neptune-like
giant planet. The physical effects presented here are of a general
nature, as such, we expect them to be relevant in a wide variety of
debris disks. In particular, we believe these results can be
qualitatively applied to the KB (although we do not pretend to present
a detailed study of the KB dynamics). A more quantitative study of the
KB or of any other specific debris disk is beyond the scope of this
letter.

\section{Simulations}

In this work we explore by means of long-term numerical simulations,
the influence of random DPs on the dynamics of cold KBLDDs. The random
DPs share physical characteristics with the ones observed in the solar
system's trans-Neptunian region, while the cold KBLDDs resemble the
observed cold population of the solar system's CKB. We constructed our
initial conditions to resemble the cold CKB because it is the
component least affected by Neptune, therefore the most stable. This
is also the most intuitive starting point for a generic statistical
study of debris disks. Among the differences with the solar system
precise conditions are: the exact quantity of DPs, a zero inclination
for our Neptune-like giant planet, and the random generated initial
conditions of the belt particles.

For our studies we employ the hybrid symplectic integrator included in
the MERCURY package \citep{Chambers99}. This integrator lets us follow
the evolution of test particles in a potential generated by several
major N-bodies plus a central star. It also permits to follow close
encounters between bodies with high accuracy by switching from a
symplectic to a Bulirsh-St\"oer integrator; the switch between
integrators takes place when particles get closer than a limit imposed
in terms of the given major body's Hill radius ($R_{\rm H} = (M_p / 3
M_{\odot})^{1/3}$).

All simulations are 1 Gyr long with an accuracy tolerance for the
Bulirsh-St\"oer integrator of $10^{-10}$, a changeover distance
between integrators of 3$R_{\rm H}$ for any major body, and a
time-step of 180 days for the symplectic integrator. The simulations
were performed on {\it Atocatl} \footnote{{\it Atocatl} is a
  supercomputer of the Instituto de Astronom\'ia at UNAM.}.

\subsection{Major Bodies}

The main central body in all simulations is a 1 $M_{\odot}$ star. 

We consider 4 different initial DP configurations: we use 10, 30, 50,
and 100 randomly generated cold DPs. The orbital parameters of all DPs
lie within the following limits: semimajor axes, $35 \,{\rm AU} <a<60
\,{\rm AU}$; eccentricities, $0.0<e<0.1$; inclinations, $0.0^{\rm
  o}<i<5.0^{\rm o}$; arguments of pericenter, $0^{\rm
  o}<\omega<360^{\rm o}$; longitudes of the ascending node, $0^{\rm
  o}<\Omega<360^{\rm o}$; and mean anomalies, $0^{\rm o}<M<360^{\rm
  o}$. DP masses take random values in the range $3.3\times10^{-6}{\rm
  M}_{\oplus}<m<2.8\times10^{-3}{\rm M}_{\oplus}$, where upper limit
corresponds to Eris's mass, while the lightest corresponds to the mass
of 2002 AW$_{197}$, this is, the biggest and one small but
significative object in our trans-Neptunian region.
 
All four DP configurations were run with and without the presence of a
giant planet. The parameters for this body were exactly the ones the
real Neptune has but with zero inclination for the sake of simplicity,
because the giant planet defines the angular momentum of the system
(i.e. this represents the natural reference system of the problem);
had we chosen different planes for the giant planet and the KBLDD an
initial rearrangement of test particles would have occurred to come
into balance with the giant planet's plane.

To better see the cumulative effect, we constructed the sets of DPs in
such a way that the larger DP sets include all the DPs of the previous
set, i.e. the set of 10 DPs is a subset of the one of 30 DPs, etc.
The total mass in DPs for 10, 30, 50, and 100 objects is respectively:
0.011, 0.032, 0.063, and 0.131 ${\rm M}_{\oplus}$; for comparison, the
CKB estimated mass is $\sim 0.01 {\rm M}_{\oplus}$
\citep{Bernstein04,FB14}.

\subsection{Test particles' initial conditions: Random Cold KBLDD}

We generate a belt of 1000 test particles that resemble the observed
cold CKB population. According to \cite{Kavelaars08}, \cite{Petit11},
and \cite{Dawson12}, the current cold CKB have orbits with semi-major
axes, $42.5\rm{AU} < a < 44.5\rm{AU}$, but mainly around 44~AU, with
inclinations, $i < 4^{\rm o}$, and eccentricities, $e< 0.05$, for most
objects of the population.

We assign the values of the orbital parameters of the particles as
follows: for $a$ we use a random Gaussian distribution with mean and
standard deviation: $\left<a\right>=44.0$ AU, $\sigma_a=1.5$ AU. For
$e$ and $\omega$ we generate a point distribution in an XY plane where
each coordinate gets random Gaussian values with mean zero and
standard deviations given by $\sigma_{(e_X,e_Y)}=0.03$; each point
represents a vector, $\vec{e}=(e_X,e_Y)$, whose magnitude,
$|\vec{e}|=\sqrt{e_X^2+e_Y^2}$, is the $e$ of the particle; also, we
define the angle between $\vec{e}$ and the X axis, $\phi_e$, as
$\omega$, therefore $\omega=\phi_e=Tan^{-1}(e_Y/e_X)$; in this manner
the initial $e$ distribution has mean and standard deviation:
$\left<e\right>=0.037$, $\sigma_e=0.019$, while $\omega$ is randomly
distributed between $0^{\rm o}$ and $360^{\rm o}$. We follow an
analogous procedure to obtain the $i$ and $\Omega$ distributions; in
this case we generate coordinates with random Gaussian points with
mean zero and standard deviations given by
$\sigma_{(i_X,i_Y)}=1.2^{\rm o}$; the resulting $i$ follows a
distribution with $\left<i\right>=1.52^{\rm o}$, $\sigma_i=0.80^{\rm
  o}$; while $\Omega$ is randomly distributed between $0^{\rm o}$ and
$360^{\rm o}$. Finally, for $M$ we use random values between $0^{\rm
  o}$ and $360^{\rm o}$.

\section{Results and Discussion}

\begin{figure}
\epsscale{1.05}
\plotone{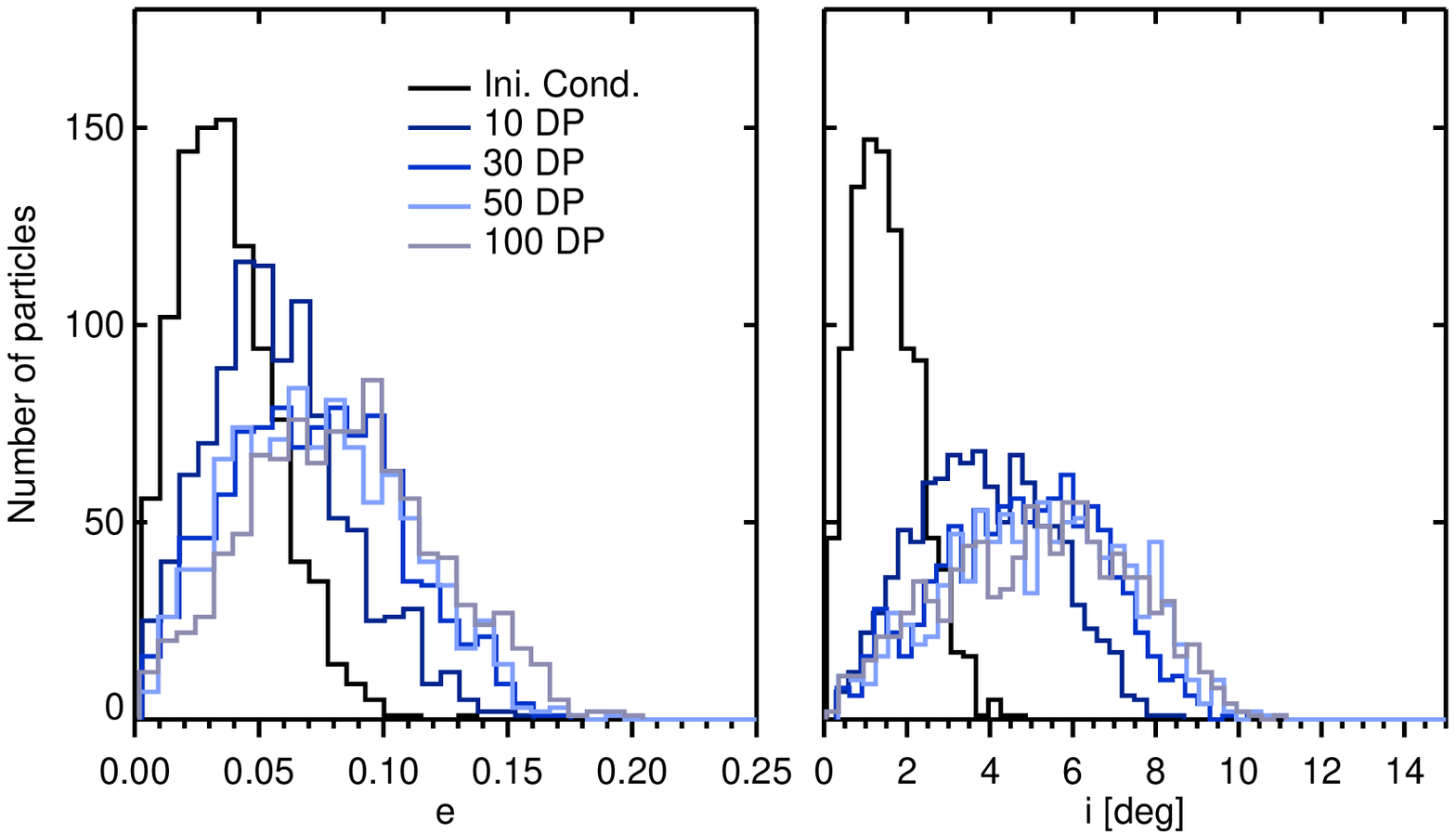}
\caption{Initial and final distributions of test particles. Left panel
  shows the initial $e$ distribution of KBLDD particles (black line)
  and the final distributions, after 1 Gyr evolution, when 10 (darker
  blue line), 30 (middle blue line), 50 (lighter blue line), and 100
  (gray line) random DPs are present in the simulation. Right panel
  is the same but for $i$.\label{his}}
\end{figure}

\begin{figure}
\epsscale{1.05}
\plotone{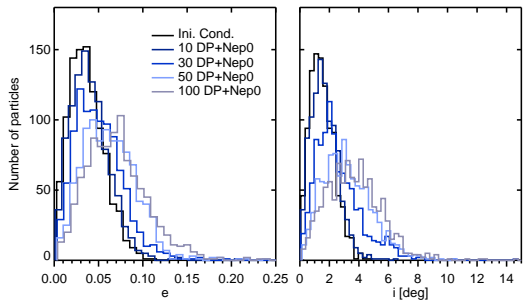}
\caption{Same as Fig.~\ref{his} but including a Neptune-like giant
  planet at 30.09 A.U. \label{hisn}}
\end{figure}

Fig.~\ref{his} shows the initial and final distributions of test
particle eccentricities (left panel) and inclinations (right panel) in
the simulations without a Neptune-like giant planet; the black line
represents the initial conditions, while the different shades of blue
represent the final distributions for 10, 30, 50, and 100
DPs. Analogously, Fig.~\ref{hisn} shows the same distributions when,
along with the DPs, a Neptune-like planet is included at 30.09
A.U. 

From Fig.~\ref{his} we see that both $e$ and $i$ shift toward larger
values as the number of DPs increases; this is to be expected as more
DPs will produce a larger number of close encounters with test
particles, resulting in larger dispersions of $e$ and $i$. A striking
difference between $e$ and $i$ distributions can be noted: while for
$e$ there are more disturbed particles as the number of DPs increases;
for $i$ there seems to exist a saturation limit, where no particles
can be heated beyond $\sim11^{\rm o}$, not even with 100 DPs, while
the mean of the distribution remains near $\sim5^{\rm o}$ with 30, 50,
and 100 DPs. The latter is result of the initial distribution of DPs;
as they are cold, with maximum initial inclinations of $5^{\rm o}$,
they do not seem to be able to push test particle's $i$ far beyond
this limit. With 10 DPs there is less dynamical heating and this limit
is not reached, remaining around $4^{\rm o}$.

\begin{figure}
\epsscale{1.05}
\plotone{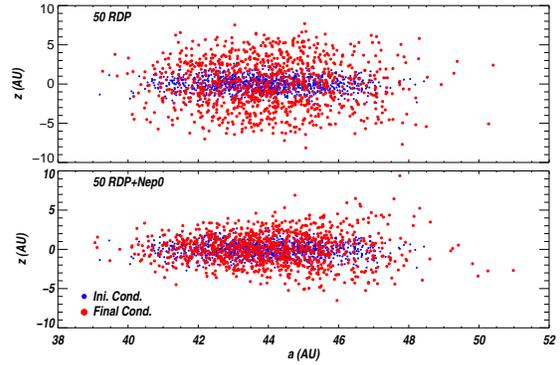}
\caption{Height over the plane ($z$) vs. semimayor axis ($a$). The
  upper panel shows the cometary nuclei initial conditions (blue) and
  final (red) for our run with 50 DPs and no giant planet. Lower panel
  shows the same but for the run with 50 DPs plus the
  giant. \label{avsz}}
\end{figure}

An interesting effect occurs when a Neptune-like planet is included in
the simulations: as seen in Fig.~\ref{hisn}, scattering induced by 10
and 30 DPs is severely damped for both $e$ and $i$
distributions. Again, with increasing DPs number, scattering of
particles becomes stronger, leading to a shift of the distributions to
higher values of $e$ and $i$. For 50 and 100 DPs, damping is slightly
less important and, although fewer in number, some particles can rise
to values of 0.20 and $11^{\rm o}$ for $e$ and $i$, respectively
(values similar to the ones reached without a giant planet). Again,
the mean values of the final $i$ distributions grow with DPs number,
but always remain below $5^{\rm o}$; even with 100 DPs, the mean is
$\sim4^{\rm o}$. This implies, contrary to intuition, that a giant
planet can act as a stabilizing agent, by helping to vertically bound
particles in its gravitational potential (see
Fig.~\ref{avsz}). Mechanisms that could be responsible for this effect
are: a) a supression on the number of close encounters of the cometary
nuclei with DPs induced by the giant planet; from our studies we find
an opposite behavior, i.e. the presence of a giant planet increases
the number of collisions due to the higher disk density produced by
its presence. b) Resonances with the giant; in this case, mean motion
resonances (MMR) in the plane of the disk produced by the giant have a
strong influence very high above the disk plane, flattening
considerably the disk; this phenomenon has been recently demonstrated
to occur in galactic disks \citep{MP15}, however the lack of
filamentary structure on Fig.~\ref{avsz}, may suggest this effect is
not important. c) Resonances induced by the DPs on the cometary
nuclei; in this case the giant planet breakes the phases of the
particle-DP interaction preventing the more efficient resonant
heating. d) A gravitational non-resonant origin based only on the
vertical force excerted by the giant; on average the giant acts like a
30 AU ring that pulls the cometary nuclei towards the plane of the
disk producing the distinctive triangle-like shape seen in
Figure~\ref{avsz}.

With enough DPs, the effect of very close encounters with DPs will be
able to overcome the stabilizing influence of the giant planet;
clearly, there must be a limit on how far this stabilizing influence
can be exerted, but in the radii we explore, we do not reach it. In
the presence of the giant, there are more close encounters due to the
higher density; this may lead to more dust production in the disk than
without the presence of the giant.

\begin{figure*}
\epsscale{2.4}
\hspace{-20mm}
\plotone{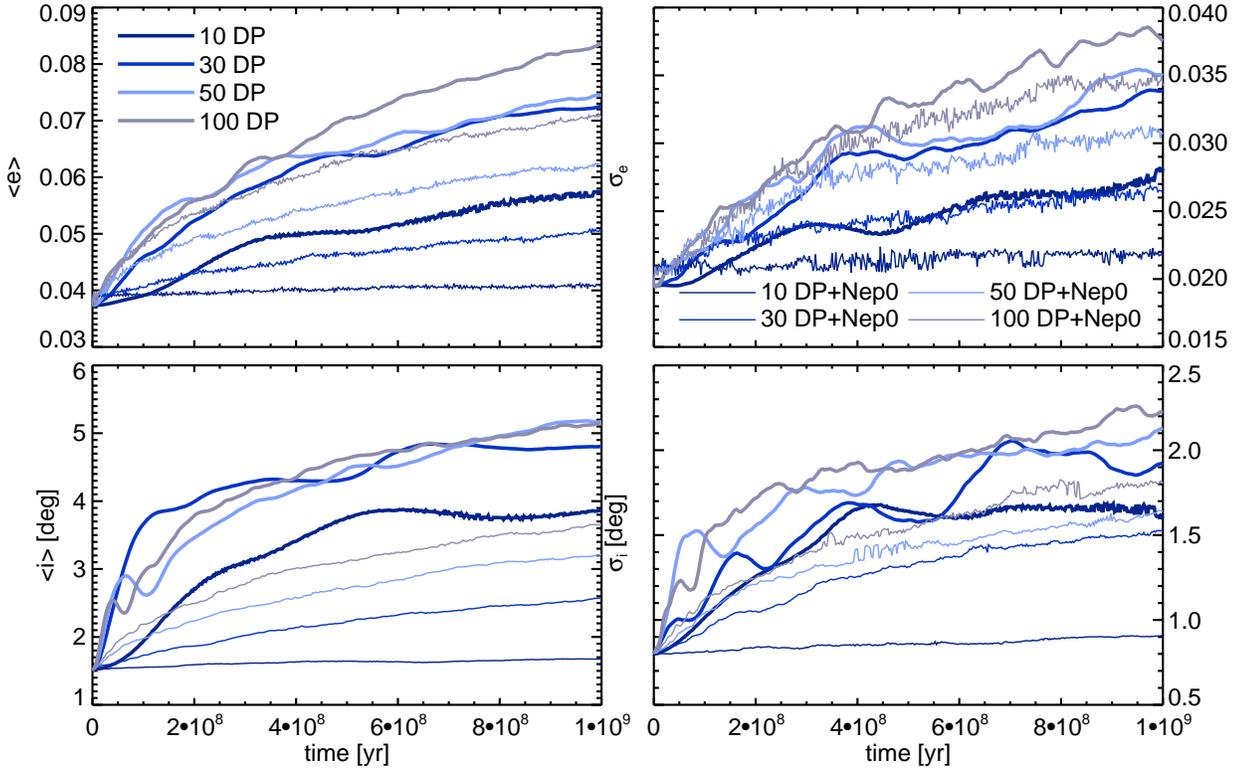}
\caption{Evolution of test particle averages and standard deviations
  for eccentricity and inclination: top-left panel shows
  $\left<e\right>$, top-right panel shows $\sigma_e$, bottom-left
  shows $\left<i\right>$, and bottom-right shows $\sigma_i$. Broad
  lines stand for simulations without a giant planet, while thin lines
  for simulations that include a Neptune-like planet. Color code is
  the same as in Fig.~\ref{his}.\label{avgsig}}
\end{figure*}

The left panels of Fig.~\ref{avgsig} show the evolution throughout the
simulations of $\left<e\right>$ and $\left<i\right>$, while the right
panels show $\sigma_e$ and $\sigma_i$, respectively. Broad lines show
the evolution produced by DPs without a giant planet, while thin lines
correspond to simulations that include a Neptune-like body. The
top-left panel of Figure~\ref{avgsig} shows how, in all 8 cases,
$\left<e\right>$ increases almost monotonically; naturally, as the
number of DPs increases, their effect on the final $\left<e\right>$
increases. The top-right panel shows a similar behavior for $\sigma_e$
(note the different scale between panels). These results strengthen
what we have seen in the previous figures: the increasing presence of
minor bodies increasingly perturbs the test particles, both with and
without a giant planet.

The growing radial heating allows test particles to encounter
resonances, replenishing them with cometary nuclei. This effect is
clear in spite of the small number of test particles we employ in our
simulations. This becomes relevant not only because of the inherently
fascinating behavior of particles trapped into resonances, but also
because it is generally assumed that, in advanced stages of debris
disks, there are no more known mechanisms able to restock the material
on resonant regions.

We also find that several of those particles are effectively trapped
by resonances with the giant planet increasing dramatically their
eccentricities. This mechanism might work as a plausible secular
process able to sustain a rate of new comets spiraling into the inner
planetary system (this rate has not been fully explained for the
KB).

By comparing the thin to the broad lines in the two bottom panels of
Fig.~\ref{avgsig}, we can see the stabilizing effect of a Neptune-like
planet: without the giant planet $\left<i\right>$ quickly grows to
reach the $5^{\rm o}$ limit found before, when a Neptune-like body is
present evolution is smooth and rising but slower; with a giant
planet, 100 DPs are required to produce a similar effect to what 10
DPs were able to achieve without the giant. Also, without a giant
planet, 30 DPs are enough to get close to some sort of saturation
point, and there is very little difference between the final values
for $\left<i\right>$ for 50 and 100 DPs; the saturation value seems to
be similar to the DPs inclination initial distribution.

A similar trend is observed in the $\sigma_i$ evolution: the maximum
dispersion reached is about $2.1^{\rm o}$ for 30, 50, and 100 DPs
without a giant planet, while with the giant this limit is about
$1.6^{\rm o}$. The effect produced by 10 bodies without the giant
planet, clearly seen in both $\left<i\right>$ and $\sigma_i$, almost
disappears in the presence of the giant planet. In our solar system
around 10 objects comparable in size to Pluto have been discovered, if
this is the total number of this kind of bodies, their effect on our
KB would be hardly noticeable; however, there is the possibility that
the total number of DPs is several times larger.

\section{Conclusions}

With the use of long-term, N-body numerical simulations we have
studied the dynamical effect of DPs on a cold debris disk
with and without the presence of a giant planet.

In the absence of the giant, DPs require only 1 Gyr to induce
substantial vertical heating on initially cold test particles; this
process increases the inclinations up to a saturation value of the
order of the highest initial DP inclinations, in our simulations,
$5^{\rm o}$. Likewise, radial heating (eccentricity dispersion)
increases rapidly, although in this case, saturation is not reached.

On the other hand, in the presence of a Neptune-like giant planet, the
contribution of the DPs to the general heating diminishes
severely. The $5^{\rm o}$ inclination limit obtained without the giant
planet is no longer reached, not even with 100 DPs; in this case, the
giant planet acts as a stability agent, concerning particle
inclinations specifically, reducing the vertical heating. Regarding
the radial heating, albeit a reduction is also observed, significant
heating remains and grows steadily in time. The gravitational
influence of the giant planet prevents the particles from dispersing,
keeping a higher density on the disk; this may have important
consequences on the rate of collisions and on dust production.

Another consequence of the heating produced by DPs is a slow but
constant secular radial migration of particles in the belt; several of
those particles are eventually trapped in the giant planet's MMRs 
where, through chaotic diffusion, they could become
part of other dynamical families \citep[e.g. Centaurs;][]{TM09}.

The continuous replenishing of resonant regions with new cometary
nuclei leads several particles through a dynamical evolution process
that produces close encounters with the giant planet. Those bodies
contribute to the influx rate of new short-period comets that may
become important from the point of view of habitability, however
observations of this mechanism are not yet available for planetary
systems other than our Kuiper belt. In the case of the solar system
this mechanism may contribute to the short-period comet influx rate,
in better accordance with observations
\citep{Emelyanenko05,Volk08,Volk13}; this is assuming the possibility
of the existence of more than ten DPs in the trans-Neptunian region.
Moreover, if the formation of several tens of DPs in the outer regions
of our solar system took place prior to the migration of Neptune, a
vertically pre-heated debris disk could have been already present when
Neptune reached its current location; such process would produce a
soft mixing between: the cold CKB population, the hot CKB population,
and the resonant objects (those swept during Neptune's migration).

\acknowledgments Authors are grateful to A. Morbidelli, R. Malhotra
and our anonymous referee
for valuable comments. We acknowledge support from UNAM-PAPIIT grants
IN-114114, IN-115413, and CONACyT grant 241732. M.A.M. thanks support
from CONACyT scholarship for conducting PhD studies.

\clearpage

\end{document}